# SAR Imaging of Moving Target based on Knowledge-aided Two-dimensional Autofocus

Xinhua Mao

Department of Electronic Engineering, Nanjing University of Aeronautics and Astronautics,

Nanjing, China, xinhua@nuaa.edu.cn

**Abstract**

Due to uncertainty on target's motion, the range cell migration (RCM) and azimuth phase error (APE) of moving targets can't be completely compensated in synthetic aperture radar (SAR) processing. Therefore, moving targets often appear two-dimensional (2-D) defocused in SAR images. In this paper, a 2-D autofocus method for refocusing defocused moving targets in SAR images is presented. The new method only requires a direct estimate of APE, while the residual 2-D phase error ( or RCM) is computed directly from the estimated APE by exploiting the analytical relationship between the 2-D phase error ( or RCM) and APE. Because the parameter estimation is performed in the reduced-dimension space by exploiting prior knowledge on phase error structure, the proposed approach offers clear advantages in both computational efficiency and estimation accuracy.

## I.  Introduction

Synthetic aperture radar (SAR) is a coherent imaging system which, by coherently processing multiple echo pulses, provides high azimuth resolution imaging of scene of interest. The coherent data processing requires accurate information on the relative geometric relationship between the radar's flight path and the target being imaged [1]. For stationary scene, this information can typically be provided by motion sensor equipped on the radar platform, e.g., GPS and/or IMU. However, for ground moving target, the geometric relationship becomes uncertainty due to the target's non-cooperative motion. Classical SAR processing assume that the illuminated area is static. It can provide accurate coherent processing for stationary scene, but not for moving targets. Therefore, moving targets in a SAR image often appear defocused [2].

The defocused effect is due to the uncompensated range cell migration (RCM) and azimuth phase error (APE) introduced by target's motion. Therefore, to refocus the moving target in a SAR image, it is necessary to estimate and correct for the residual RCM and APE. The estimation of APE is relatively easy. Typically, it can be implemented using conventional autofocus methods [3-5]. For RCM correction, keystone transform (KT) [6-7] is well known for its capability of eliminating an arbitrary linear range migration without the kinetic information of moving target. The main drawback of KT is that it ignores the high order range migrations. In more general case, for example, the target moves with an arbitrary path, ISAR processing is proposed to refocus the moving target [8-9]. However, in the image domain, the residual 2-D phase error may contain not only the APE and RCM terms, but also high order terms in range frequency (This can be clearly seen from the analysis in Section II). In order to apply standard ISAR processing, the defocused SAR subimage of moving target has to be mapped into data domain, i.e., an inversion mapping operation corresponding to the algorithm used to form the SAR image should be performed on the subimage data. This inversion mapping will increase the computational complexity, and also introduce addition interpolation errors [9].

In this paper, we propose a knowledge-aided two-dimensional autofocus approach to refocus the moving target directly in the SAR image domain. As the basis of the proposed method, we first investigate the residual 2-D phase error structure of moving target in polar format algorithm processing framework. Then by exploiting the derived a priori knowledge on phase error structure, the residual 2-D phase error (or the residual RCM if range defocus effect can be ignored) can be mapped directly from the APE. That is to say, for the proposed 2-D refocus method, only a direct estimation of APE is required.

## II. Two-dimensional Phase Error Model

*A. Signal Model in Phase History Domain*

Consider a spotlight-mode SAR operating with the geometry depicted by Fig.1. For notational simplification, we assume that the radar operates in broadside mode, the results are easily generalized to squint mode. Let $t$ represent the slow time. The distance between the antenna phase center (APC) and the scene center (Point O) is $\mathbf{r_c} \equiv r_c(t)$, which along with the instantaneous squint angle $\mathbf{\theta} \equiv \theta(t)$ and the incidence angle $\mathbf{\varphi} \equiv \varphi(t)$ determines the instantaneous coordinate $(\mathbf{x_a}, \mathbf{y_a}, \mathbf{z_a}) \equiv [x_a(t), y_a(t), z_a(t)]$ of the APC. Note that the bold face variables in this paper are all functions of slow time $t$. At the aperture center ($t=0$), the incident angle is denoted as $\varphi_{ref}$ and the azimuth angle $\theta = 0$. Without loss of generality, a moving target in the ground plane with an arbitrary motion trajectory is assumed.

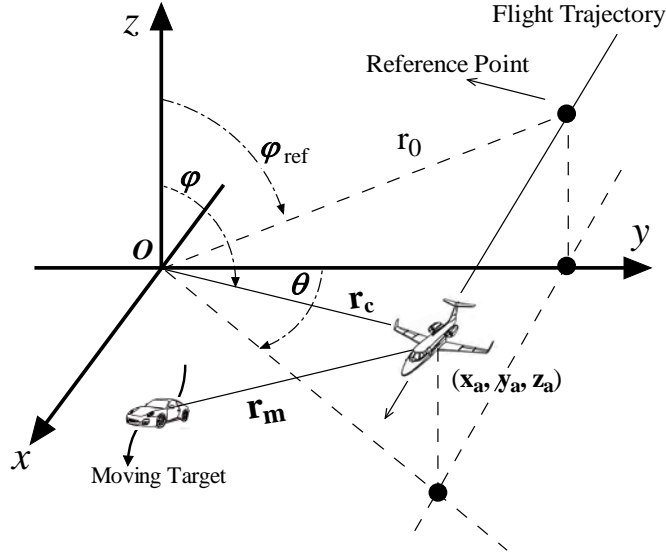

Fig. 1. Spotlight SAR data collection geometry

Assuming the radar emitted a wideband signal, for the moving target, the 2-D echo signal after demodulation and match filtering can be expressed as

$$S(t, f_r) = A \cdot \exp\left\{-j\frac{4\pi}{c}(f_c + f_r)r_m(t)\right\}, \qquad (1)$$

where $t$ is the azimuth time, $f_r$ is the range frequency, $f_c$ is the carrier frequency, $c$ is the speed of propagation, $r_m(t)$ represents the instantaneous range from the moving target to radar, $A$ includes all the nonessential amplitude factors.

From (1), we can see that in the phase history domain the phase include two terms, one is the azimuth phase term which is range-frequency independent, the other one is the range migration term. It is clear that the two terms are linear related.

*B. Residual 2-D Phase Error after PFA Processing*

In SAR-based moving target detection and imaging system, SAR image formation algorithm is often first applied to produce a complex image of the whole scene. In the SAR image, stationary targets are well focused, while moving targets appear defocused due to uncompensated 2-D phase error. To correct for this phase error and refocus moving targets, it is desirable to derive the 2-D phase error model for moving target in SAR image. To this end, in the following, we will analyze the 2-D phase error in the framework of polar format algorithm.

To proceed with the PFA, the echoes must be motion compensated to a reference point. Generally, the scene center is selected as the reference. In the previous subsection, we have let $r_c(t)$ represent the instantaneous range from the radar to this stationary reference point, then after motion compensation, the signal becomes

$$S_M(t, f_r) = A \cdot \exp\left\{ j\frac{4\pi}{c}(f_c + f_r)[r_c(t) - r_m(t)] \right\}. \tag{2}$$

In [10], we provide a new interpretation of polar reformatting, where the range resampling is considered as a range frequency scaling transformation, and the azimuth resampling is interpreted as a combination of RCM linearization and the Keystone transform (KT).

The range frequency scaling transform has a scaling factor of $\delta_r = \sin\varphi_{ref} / (\sin\varphi \cos\theta)$

and an offset of $f_c(\delta_r - 1)$. Therefore, after range resampling, the signal in (1) becomes

$$S_R(t, f_r) = S_M\left[t, \delta_r f_r + f_c(\delta_r - 1)\right] \\ = A \cdot \exp\left\{j\frac{4\pi}{c}(f_c + f_r)\sin\varphi_{ref}\varpi(t)\right\}, \quad (3)$$

where $\varpi(t) = [r_c(t) - r_m(t)]/(\sin\varphi\cos\theta)$.

The second step of PFA is azimuth resampling. We divide it into two cascaded resampling procedures, i.e., RCM linearization and KT. RCM linearization is a azimuth time transformation, which is independent of the range frequency, to linearize $\tan\theta$. Mathematically, this procedure can be implemented by performing a change-of-variable on the azimuth time, denoted as $t \to \vartheta_a(t)$. Therefore, after RCM linearization, the signal in (3) becomes

$$S_{A1}(t, f_r) = S_R[\vartheta_a(t), f_r] = A \cdot \exp\left\{j\frac{4\pi(f_c + f_r)\sin\varphi_{ref}}{c}\eta(t)\right\}, \quad (4)$$

where $\eta(t) = \varpi[\vartheta_a(t)]$ is a function with respect to azimuth time $t$, it can be expressed by Taylor series expansion as following

$$\eta(t) = a_0 + a_1 t + \xi(t) \quad (5)$$

where $a_0$ is the constant term, $a_1$ is the coefficient of linear term, $\xi(t)$ includes the quadratic and higher order terms.

The final step of polar reformatting is to perform the KT on (4), which results in

$$S_{A2}(t, f_r) = S_{A1}\left[\frac{f_c}{f_c + f_r}t, f_r\right] \\ = A \cdot \exp\left\{j\frac{4\pi\sin\varphi_{ref}}{c}\left[a_0(f_c + f_r) + a_1 f_c t + (f_c + f_r)\xi\left(\frac{f_c}{f_c + f_r}t\right)\right]\right\}, \quad (6)$$

To be consistent with traditional symbol expression, we can define

$$X = \frac{4\pi\sin\varphi_{ref}f_c}{c}t \\ Y = \frac{4\pi\sin\varphi_{ref}}{c}(f_c + f_r) \quad (7)$$

as the spatial frequency with respect to $x$ and $y$ direction, respectively. Note that the range frequency has an offset $Y_0 = 4\pi \sin \varphi_{ref} f_c / c$. After these change-of-variables, (6) can be rewritten as

$$S_{A2}(X, Y) = A \cdot \exp\left\{j\left[a_0 Y + a_1 X + Y\xi\left(\frac{X}{Y}\right)\right]\right\} \tag{8}$$

In (8), the linear phase terms with respect to the range and azimuth frequency are the basic imaging terms whose coefficients indicate the position of moving target in SAR image; The coupling term

$$\Phi_e(X,Y) = Y\xi\left(\frac{X}{Y}\right) \tag{9}$$

is the undesirable 2-D phase error term which must be corrected in order to focus the moving target.

C. *Analytical Structure of Residual 2-D Phase Error*

To show the analytical structure of 2-D phase error, Taylor series expansion of phase error with respect to range frequency is performed on (9), which results in

$$\Phi_e(X, Y) = \phi_0(X) + \phi_1(X)(Y - Y_0) + \phi_2(X)(Y - Y_0)^2 + \cdots, \tag{10}$$

where

$$\phi_0(X) = Y_0 \xi\left(\frac{X}{Y_0}\right), \tag{11}$$

$$\phi_1(X) = \xi\left(\frac{X}{Y_0}\right) - \frac{X}{Y_0}\xi'\left(\frac{X}{Y_0}\right), \tag{12}$$

$$\phi_2(X) = X^2 \xi''\left(\frac{X}{Y_0}\right) / (2Y_0^3), \tag{13}$$

where $\xi'(X/Y_0)$ and $\xi''(X/Y_0)$ respectively denote the first and second derivatives of $\xi(X/Y_0)$. In (10), the $\phi_0(X)$ term corresponds to the APE, the $\phi_1(X)$ term to the residual RCM, and the $\phi_2(X)$ term and other high-order terms are related to the range defocus.

Consulting (11) and (12), it is easy to get the analytical relationship between $\phi_0(X)$ and $\phi_1(X)$:

$$\phi_1(X) = \frac{1}{Y_0}\left\{\phi_0(X) - X \cdot \frac{\mathrm{d}\phi_0(X)}{\mathrm{d}X}\right\}. \tag{14}$$

Consulting (9) and (11), we can also get the analytical relationship between $\phi_e(X,Y)$ and $\phi_0(X)$:

$$\phi_e(X,Y) = \frac{Y}{Y_0}\phi_0\left(\frac{Y_0}{Y}X\right) \tag{15}$$

### III. Refocus of Moving Target by 2-D Autofocus

To refocus the moving target in PFA image, it is necessary to estimate and correct for the 2-D coupling error terms in (8). If we have no a priori information on the 2-D phase error, it is natural to estimate the 2-D phase error in a blind manner, that is, estimate the 2-D phase error directly or estimate APE and RCM using ISAR method when range high order terms can be ignored. Due to the high dimensionality of the error parameters and limited number of training samples, these blind estimation approaches suffer from not only high computational complexity, but also poor estimate accuracy. Fortunately, from the analysis in the previous section, we have known that the residual 2-D phase error after PFA processing is not totally unknown. In fact, the residual 2-D phase error has a specific structure. From (15), we can see that the 2-D phase error can be determined by the azimuth phase error. By incorporating this a priori knowledge, the estimation of residual 2-D phase error can be done in a reduced-dimensional parameter space. That is to say, we need only estimate APE directly, while the 2-D phase error can then be computed from APE by exploiting the analytical relationship between 2-D phase error and APE shown in (15). Based on this idea, a moving target imaging approach based on knowledge-aided 2-D autofocus is proposed, whose processing flow is shown in Fig. 2. The key to this algorithm consists of two parts: estimation of the APE and computation of the 2-D phase error.

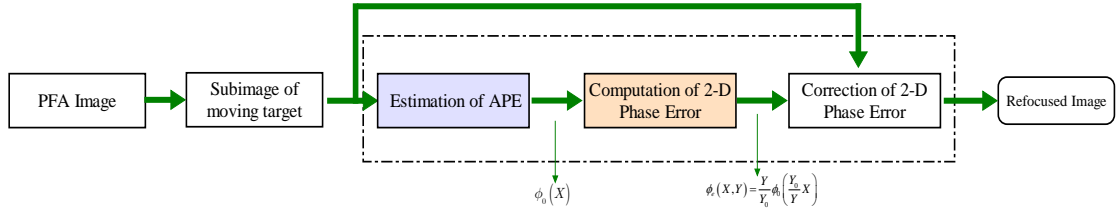

Fig. 2.    Flowchart of the proposed moving target imaging approach

*Azimuth Phase Error Estimation*

The APE estimation can be accomplished by conventional autofocus algorithms. If these algorithms are exploited without proper modifications, however, the phase error estimation performance degrades in the case that the residual range migration cannot be ignored. A straightforward method to resolve this problem is to reduce the range resolution prior to the estimation of the phase error. As such, the residual range migration is confined into one range resolution cell. The APE is then estimated using conventional autofocus algorithms.

*Computation of 2-D Phase Error*

According to (14), we can see that the mapping from APE to the 2-D phase error is one-to-one and flight-path independent. Thereby, once the APE is estimated, the 2-D phase error can be calculated directly. This direct computation of 2-D phase error eliminates a 2-D blind estimation process, thereby possessing a high computational efficiency.

### IV. Experimental Results

Real data collected by an airborne spotlight SAR system is applied to demonstrate the effectiveness of the proposed method. Fig.3 is a SAR image produced by PFA processing, where the stationary scene are well focused, but a moving target marked with a white square in the figure appears 2-D smeared. A subimage entirely containing the defocused moving target is selected as the input for the proposed autofocus method. Fig.4 (a) and (b) show this subimage and its corresponding range compressed image, respectively. From the range compressed image, we can clearly see the residual RCM effect.

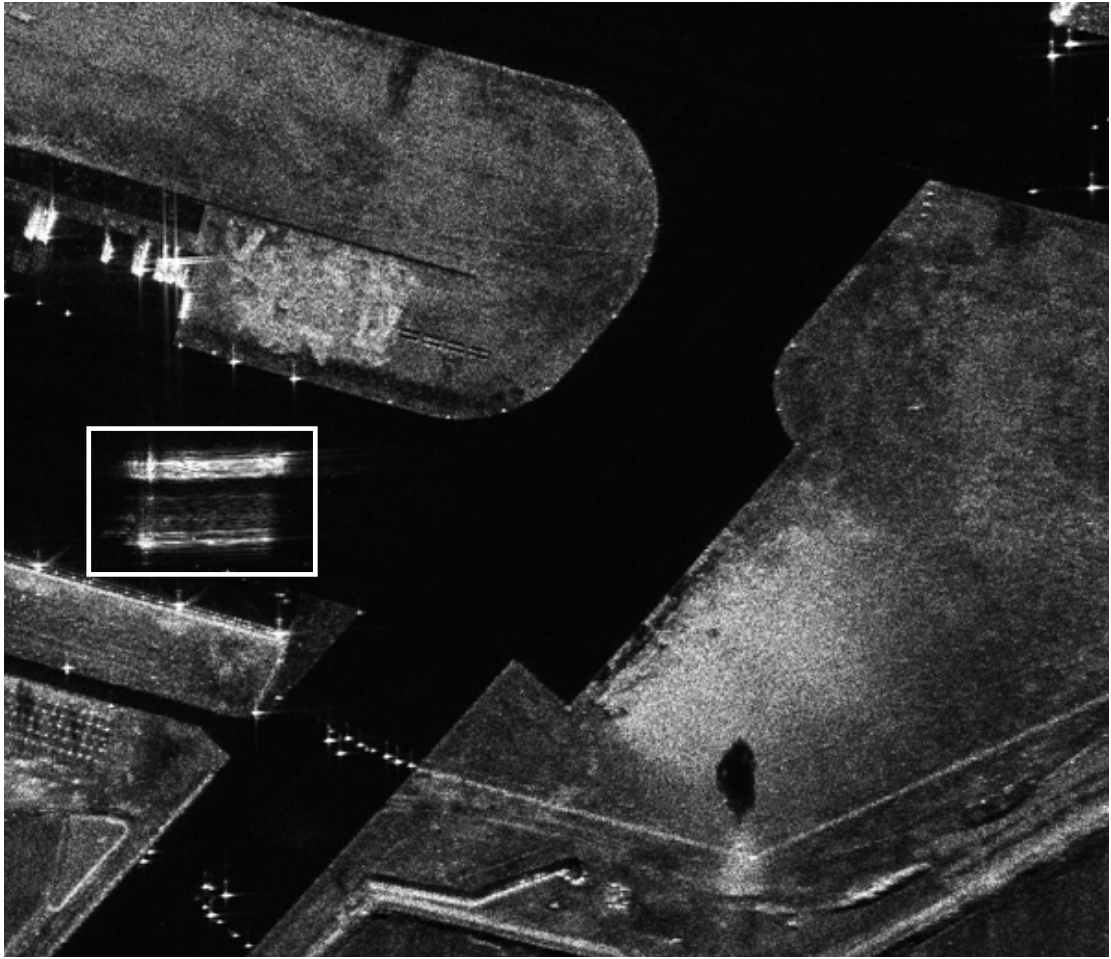

Fig. 3.　SAR imagery produced by PFA

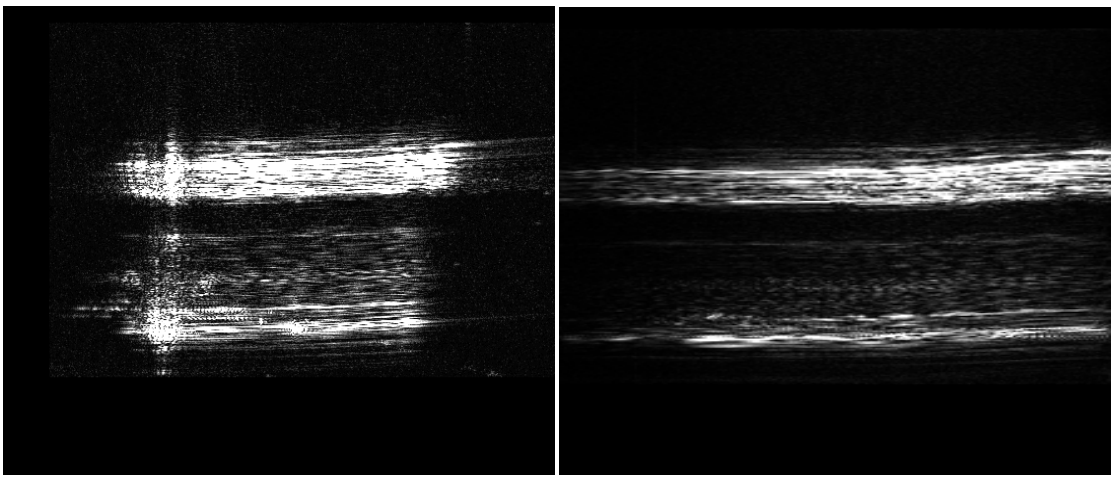

(a)　　　　　　　　　　　　　　　　　　(b)

Fig. 4.　(a) Subimage of moving target and (b) its range compressed image

The proposed 2-D autofocus method is applied to the selected subimage data to refocus the moving target. Fig.5 (a) and (b) give the refocused subimage and its corresponding range compressed image, respectively. Obviously, the 2-D defocus effect is perfectly eliminated, the target seems well focused.

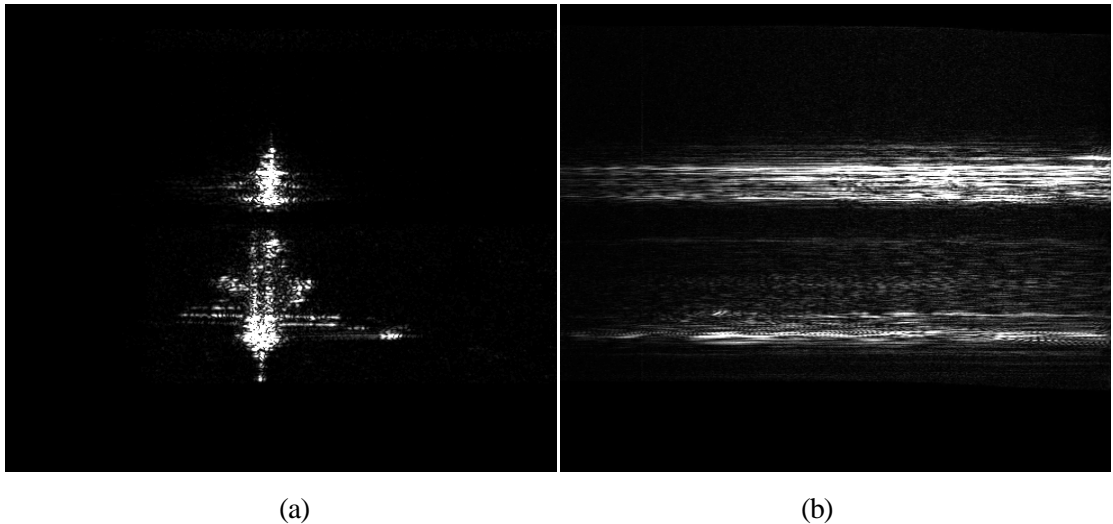

(a) (b)

Fig. 5. (a) Subimage of moving target and (b) its range compressed image after the proposed 2-D autofocus processing

### V. Conclusion

In this paper, we revealed the analytical structure of residual 2-D phase error for moving target in SAR imagery. By incorporating the derived a priori knowledge on the phase error structure, we then proposed a knowledge-aided 2-D autofocus algorithm to refocus the moving target in SAR imagery. Because the refocus process is performed directly in the SAR image domain, and parameter estimation is done in the reduced-dimension space, the proposed approach offers clear advantages in both computational efficiency and estimation accuracy as compared with the ISAR based algorithms.

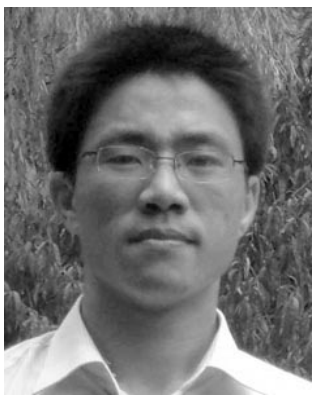

**Xinhua Mao** was born in Lianyuan, China, in 1979. He received the B.S. and Ph.D. degrees from the Nanjing University of Aeronautics and Astronautics (NUAA), Nanjing, China, in 2003 and 2009, respectively, all in electronic engineering.

Since 2009, he joined the Department of Electronic Engineering, NUAA, where he is now an associate professor. His research interests include radar imaging, and ground moving target indication (GMTI).